\documentclass[twocolumn]{aastex63}
\usepackage{bm}
\shorttitle{The pulsar gamma-ray emission}
\shortauthors{Cao \& Yang}
\begin{document}
\title{The energy-dependent gamma-ray light curves and spectra of the Vela pulsar in the dissipative magnetospheres}
\author{Gang Cao}
\author{Xiongbang Yang}
\affiliation{Department of  Mathematics, Yunnan University of Finance and Economics, Kunming 650221, Yunnan, P. R. China; gcao@ynufe.edu.cn,xbyang@ynufe.edu.cn}
\begin{abstract}
We study the  pulsar energy-dependent $\gamma$-ray light curves and  spectra from curvature radiation in the dissipative magnetospheres. The dissipative magnetospheres with the combined force-free (FFE) and Aristotelian (AE) are computed by a pseudo-spectral method with the high-resolution simulation in the rotating coordinate system, which produces a near force-free field structure with the dissipative region only near the equatorial current sheet outside the light cylinder (LC). We use the test particle trajectory method to compute the energy-dependent $\gamma$-ray light curves, phase-average and phase-resolved spectra by including both the  accelerating electric field and radiation reaction. The predicted energy-dependent $\gamma$-ray light curves and spectra are then compared  with those of the Vela pulsar observed by Fermi. Our results can generally reproduce the  observed trends of the energy-dependent $\gamma$-ray light curves and  spectra for the Vela pulsar.
\end{abstract}


\section{Introduction}
Pulsars are  rapidly rotating  neutron star with extremely strong magnetic field, which can convert a substantial fraction of rotational kinetic energy into particle acceleration and radiation.
These objects can emit the broadband electromagnetic emission from the radio to very high-energy GeV, and even TeV bands. The radiation from pulsars originates from the high-energy particles accelerated by unscreened electric fields in the magnetosphere. These high-energy particles flow out along the open field lines and then produce the broadband  electromagnetic spectra  by the synchrotron, curvature, and inverse-Compton processes. The Fermi Gamma-Ray Space Telescope launched in 2008 has revolutionized  our knowledge of pulsar physics with the detection of more than 270 $\gamma$-ray pulsars. The Second Fermi Pulsar Catalog (2PC) contained 117 $\gamma$-ray pulsars with the measurement of good-quality light curves and spectra\citep{abd10,abd13}. They are divided into young radio-loud, young radio-quiet, and millisecond pulsars. Fermi  pulsars usually show the double-peak $\gamma$-ray light curves with $\sim0.5$ peak separation and the first peak typically lags behind the radio peak. An apparent feature is that the relative ratio of first to second $\gamma$-ray peaks decrease as the the photon energy increases for Vela and Geminga. The observed $\gamma$-ray spectra have the exponentially cutoff power-law shape with the cutoff energy in the range of 1-5 GeV. Good-quality $\gamma$-ray data from Fermi observation provide a wealth of information to explore the radiation mechanism and particle acceleration in the magnetosphere. However, it is still unclear where the particles in the magnetosphere are accelerated and how the emitting photons are produced. The particle acceleration and pulsed emission in the magnetosphere is connected to  the structure of the global pulsar magnetosphere. It is compulsory to self-consistently solve for Maxwell equations and particle dynamics and radiation to model the global pulsar magnetosphere and explain the observed pulsed emission.

Significant progresses towards the self-consistent modeling of global pulsar magnetosphere have been made over the last decades. It is expected that the pulsar magnetospheres are filled
with plasmas by the pair production\citep{gol69}. A good first guess about plasma-filled magnetosphere is referred to as  the force-free electrodynamic, which corresponds to the zeroth order approximation of global pulsar magnetosphere.
Force-free solutions have recently become available by the numerical simulation pioneered by \citet{con99} for the aligned rotator and then by \citet{spi06} for the oblique rotator. The force-free solutions are later studied by the time-dependent simulations with the spectral method \citep{par12,pet12,cao16a,pet16} and  the finite-difference method \citep{kal09,con10}. All these force-free simulations reveal the existence of an equatorial current sheet outside the LC, which is suggested to be  an alternative site of the pulsed emission in addition to the ``gap" regions \citep{bai10,har15,bra15,bog18,har18}. The force-free solutions are thought to be  closest to realistic pulsar magnetospheres, but no particle acceleration and  radiation is allowed in the magnetosphere. Therefore, some dissipation mechanism  should be included to address the problem of particle acceleration and radiation production in the magnetosphere. The resistive magnetosphere by introducing a conductivity parameter  are then developed to self-consistently include the dissipative regions in the magnetosphere, which can switch the magnetospheric solutions from the vacuum to force-free limits \citep{li12,kal12a,cao16b}. The resistive solutions have been also used to model the pulsar light curves and spectra by including the accelerating electric field from the simulations \citep{kal12b,kal14,kal17,cao19,yang21}. However, it is not clear about the  origin and physical motivation of these resistivities by involving an arbitrary conductivity. The aforementioned fluid description offer us a good knowledge of global pulsar magnetosphere with the macroscopic current and charge densities. However, the fluid methods can not model the source of particles that provide the current and charge densities. Therefore, a full self-consistent PIC (Particle-In-Cell) method is developed to model the global pulsar magnetosphere by including the particle dynamics and radiation reactions \citep{phi14,che14,phi15,bel15,cer15,kal18,bra18}. The PIC methods are also connected to the observational signature by simultaneously extracting the pulsar light curves from the simulations \citep{cer16,phi18,kal18}. However, the  Lorentz factors from the PIC simulation are  much smaller than ones in the real pulsars, which is  not enough to explain the observed $\gamma$-ray photons.

Aristotelian electrodynamics  is a good approximation between the resistive model and PIC model, which can  include the back-reaction of emission onto particle dynamics and allow for a local dissipation where the force-free condition is violated. The AE method has been introduced to model the structure of pulsar magnetospheres \citep{gru12,gru13,con16}. A clever method by combining FFE and AE is also proposed to model the dissipative pulsar magnetosphere for the aligned rotator \citep{pet20a} and for oblique rotators \citep{cao20,pet22}. Recently, \citet{cao22} perform the high-resolution simulations of the combined FFE and AE magnetospheres to better capture the dissipative regions near the current sheet outside the LC.   The pulsar light curves and spectra are also computed by using the  accelerating electric field from the simulations.  We confirm that the the particle acceleration and the $\gamma$-ray radiation is produced near the current sheets outside the LC.  Moreover,  the predicted light curves and spectra can general reproduce the  double-peak $\gamma$-ray profiles and GeV spectral cutoff energy in agreement with the Fermi observations for  the high pair multiplicity. However, the light curves and spectra prediction are not  directly compared with the  Fermi observed $\gamma$-ray data in this study. In this paper, we compute the energy-dependent light curves, phase-average and phase-resolved spectra by expanding the study of \citet{cao22} and directly compare the predicted energy-dependent light curves and spectra with the Fermi observation for the Vela pulsar.

\begin{figure}
\epsscale{1.0}
\plotone{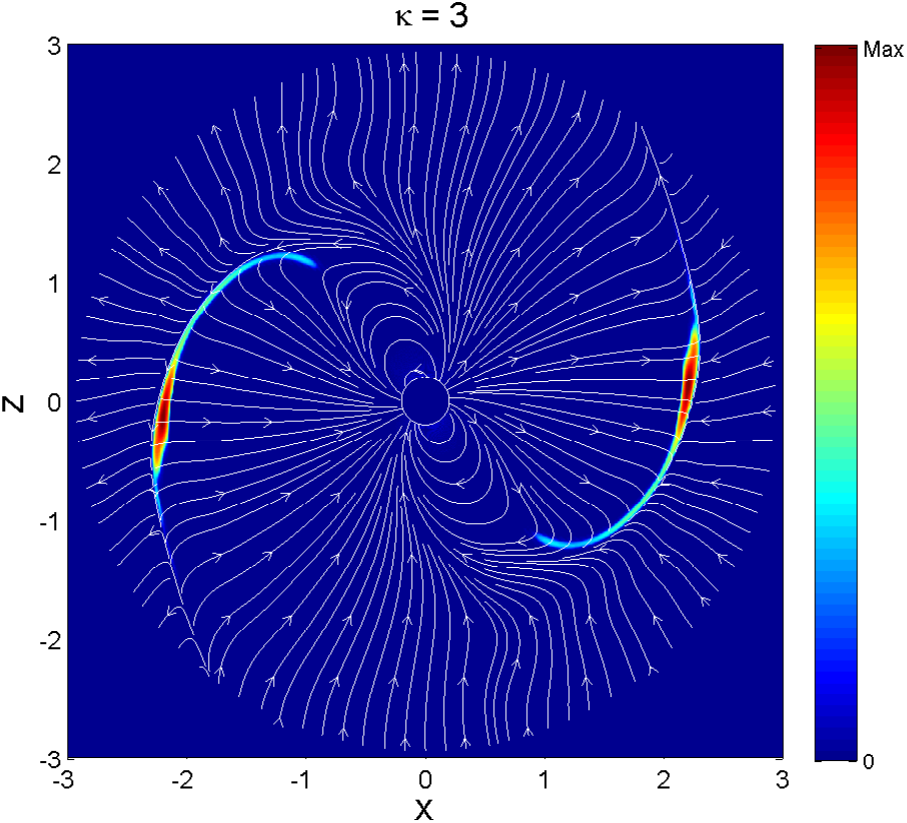}
\caption{The structure of magnetic field line and the distribution of the accelerating electric field  for a  $65^\circ$ rotator  with the pair multiplicity $\kappa=3$ in the $x$$-$$z$ plane. }
\label{Fig1}
\end{figure}

\begin{figure*}
\center
\begin{tabular}{cccccccc}
\\
\includegraphics[width=13 cm,height=6. cm]{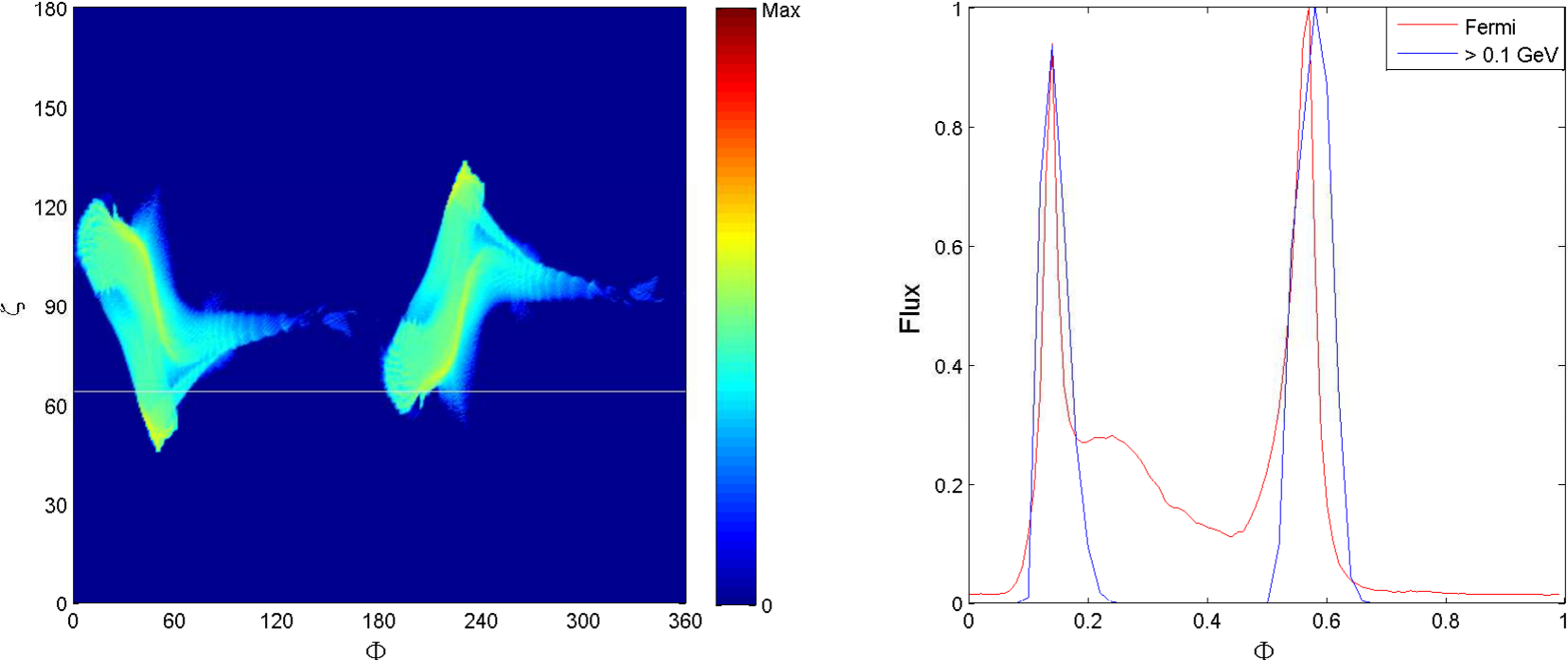}
\end{tabular}
\caption{The sky maps and the  $\gamma$-ray light curves at $>$ 0.1 GeV energies for the Vela pulsar. The model parameters are  $B_{*}=4\times10^{12} \, \rm G$, $P=0.089 \, \rm s$, $\kappa=17$, $\alpha=65^\circ$ and $\zeta=64^\circ$. The blue curve is the predicted $\gamma$-ray light curve, the red curve is the Fermi observed  data taken from \citet{abd13}. \\}
\label{Fig2}
\end{figure*}

\begin{figure*}
\center
\begin{tabular}{cccc}
\includegraphics[width=19 cm,height=5.5 cm]{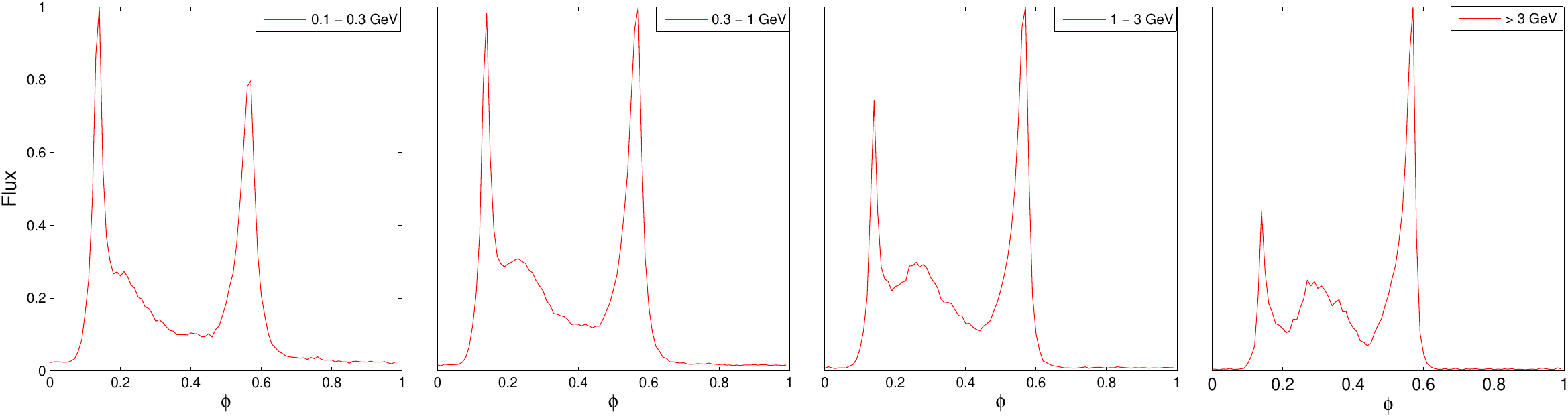}\\
\includegraphics[width=19 cm,height=5.5 cm]{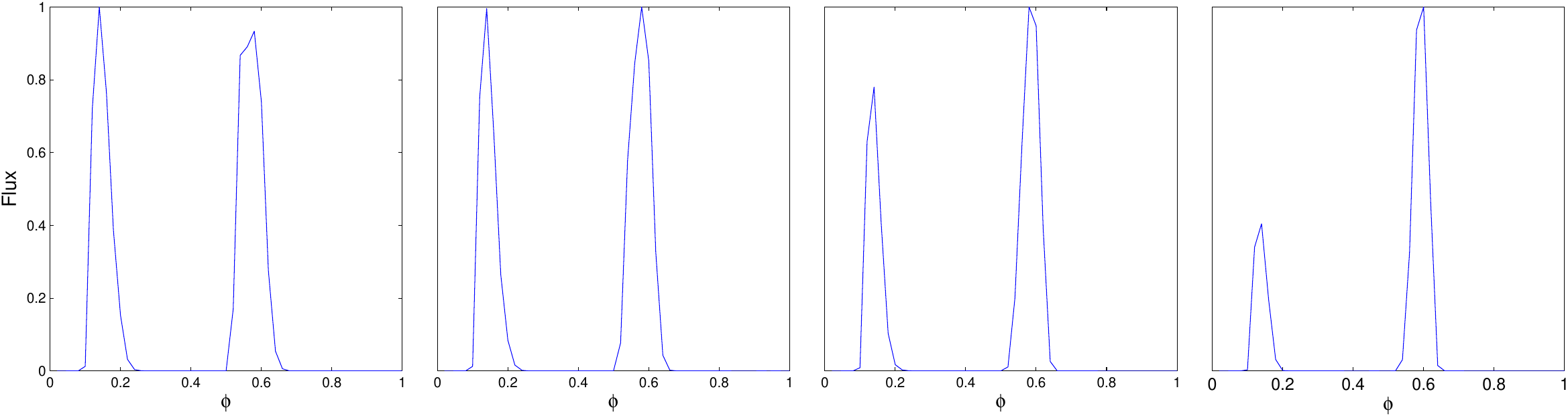}
\end{tabular}
\caption{A comparison of the predicted energy-dependent $\gamma$-ray light curves and the Fermi observed ones for the Vela pulsar.  The top panel: the Fermi observed $\gamma$-ray light curves taken from \citet{abd13}. The bottom panel: the predicted energy-dependent $\gamma$-ray light curves  with the same model parameters as in Figure \ref{Fig2}. \\}
\label{Fig3}
\end{figure*}

\section{dissipative pulsar magnetospheres}
The structure of pulsar magnetospheres are obtained by solving the time-dependent Maxwell equations in  the rotating coordinate system \citep{mus05,pet20b}
\begin{eqnarray}
{\partial {\bf B}\over \partial t'}=-{\bf \nabla} \times ({\bf E}+{\bm V}_{\rm rot}\times{\bf B}) ,\\
{\partial  {\bf E}\over \partial t'}={\bf \nabla} \times ({\bf B}-{\bm V}_{\rm rot}\times{\bf E})-{\bf J}+{\bm V}_{\rm rot}\nabla\cdot{\bf E},\\
\nabla\cdot{\bf B}=0\;,\\
\nabla\cdot{\bf E}=\rho_{\rm e}\;,
\end{eqnarray}
with the 3D spectral method \citep{cao22}, where  ${\bf J}$  is the current density, $\rho_{\rm e}$ is the charge density, ${\bm V}_{\rm rot}={\bf \Omega } \times {\bf r}$ is the corotating velocity. The current density with the combined FFE and AE is implemented to the  Maxwell equations by introducing the pair multiplicity  $\kappa$  \citep{cao20}
\begin{eqnarray}
{\bf J }=  \rho_e  \frac {{\bf E} \times {\bf B}}{B^2+E^2_{0}}+ (1+\kappa)\left|\rho_e\right| \frac{ (B_0{\bf {B}}+E_0{\bf {E}}) }{ B^2+E^2_{0}}.
\label{Eq6}
\end{eqnarray}
where  $B_0$ and $E_0$ are the magnetic and electric field in the frame in which ${\bf E}$ and ${\bf B}$ are parallel. $E_0$  is denoted as $E_{\rm acc}$ in the rest of the paper, which represents the effective accelerating electric component. $|\rho_e|$ can be interpreted as the charge density of the primary electrons, $\kappa|\,\rho_e|$ can be interpreted as  the  charge density of the secondary pairs from the pair cascades. Therefore, the pair multiplicity $\kappa$ is  physically related with the pair cascade processes in the magnetosphere. The quantities $B_0$ and $E_0$ is defined by the Lorentz invariants
relations
\begin{eqnarray}
B^2_{0}-E^2_{0}={\bf B}^2-{\bf E}^2, \,\, E_{0}B_{0}={\bf E}\cdot {\bf B},  \,\, E_{0}\geq0.
\end{eqnarray}

\begin{figure*}
\center
\begin{tabular}{c}
\includegraphics[width=6.5 cm,height=6. cm]{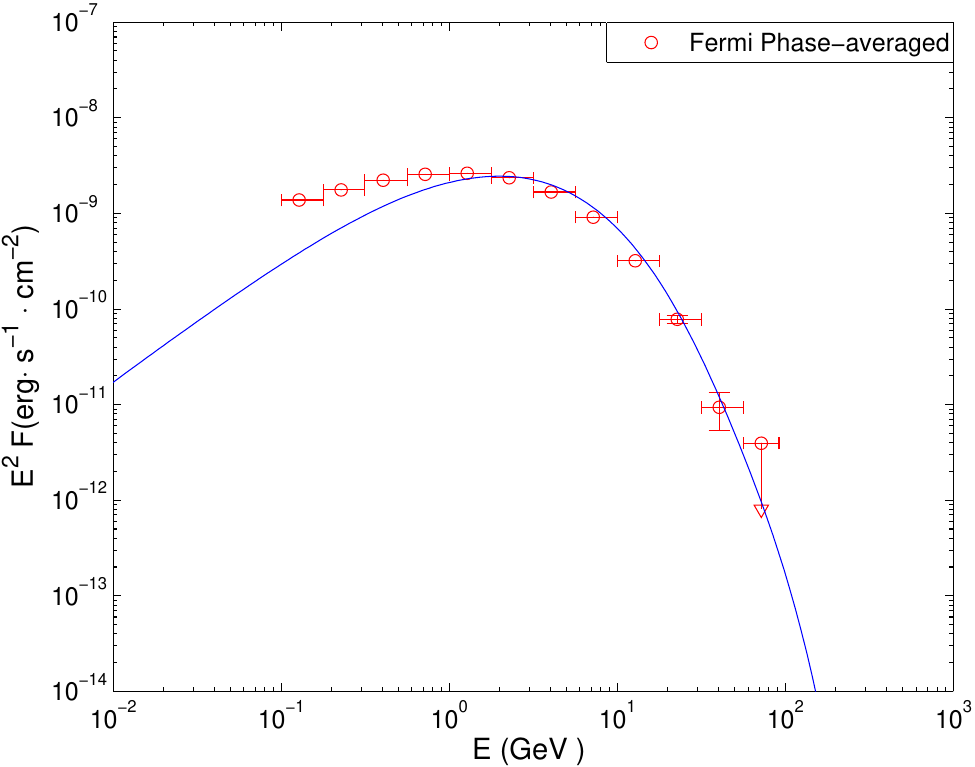} \qquad
\includegraphics[width=6.5 cm,height=6. cm]{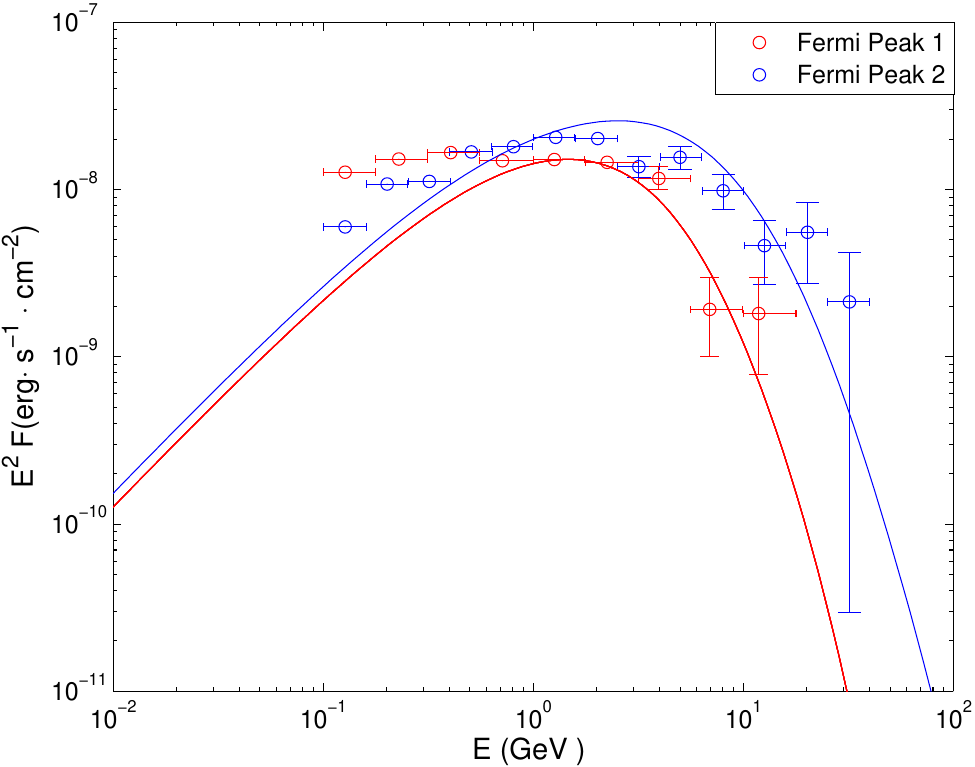} \qquad
\end{tabular}
\caption{The predicted $\gamma$-ray phase-averaged spectra and the phase-resolved spectra for the Vela pulsar with the same model parameters as in Figure \ref{Fig2}. The Fermi observed data is taken from \citet{abd13}. \\ }
\label{Fig4}
\end{figure*}

The magnetic field is initialized as  an  oblique vacuum dipole  magnetic field with magnetic inclination angles $\alpha$  from $0^\circ$ to $90^\circ$ with a interval of $5^\circ$.
The inner boundary condition at the stellar surface is enforced  with a rotating electric field ${\bf {E}} = -( {\bf \Omega } \times {\bf r} ) \times {\bf B}$.
A non-reflecting boundary condition is implemented  to prevent spurious  reflection from the outer boundary.  A high resolution with $N_r \times N_{\theta} \times N_{\phi}=129 \times 64 \times 128$ is used to obtain the accurate magnetospheric solutions from  the stellar  surface  $r=0.2 $ $r_{\rm L}$ to $r=3 $ $r_{\rm L}$. The pair multiplicity is set to $\kappa=\{0,1,3\}$.
The dissipative pulsar magnetospheres with the combined FFE and AE  is computed by applying the force-free description where $E \leq B$ and the AE  description where $E > B$.  It is expected that our model will  locate the dissipative region at the current sheet outside the LC where $E > B$.

Figure \ref{Fig1}  shows the structure of magnetic field line and the distribution of the accelerating electric field $E_{\rm acc}$ for a  $65^\circ$ rotator with the pair multiplicity $\kappa=3$ in the $x$$-$$z$ plane. We see that the combined FFE and AE magnetosphere tends to the force-free solution  and the dissipative region is only confined to  near the current sheet outside the LC. For the higher $\kappa$ values, we can expect that the  magnetospheric solutions keep the near  force-free structures  and the accelerating electric fields also decrease as the pair multiplicity $\kappa$ increases (see \citealt{cao22}). The numerical integration for the higher $\kappa$ value become cumbersome because of the stiff nature of the pair multiplicity term. Therefore, the $E_{\rm acc}$ value at $\kappa>3$  is determined by the scaling  relation
\begin{eqnarray}
E_{\rm acc}=E_{\rm acc,0}\frac{\kappa_0}{\kappa}\qquad(\kappa>\kappa_0)
\end{eqnarray}
where $E_{\rm acc,0}$ is the accelerating electric field at $\kappa_0=3$.

\begin{figure*}
\center
\begin{tabular}{cccccccc}
\\
\includegraphics[width=13 cm,height=6.5 cm]{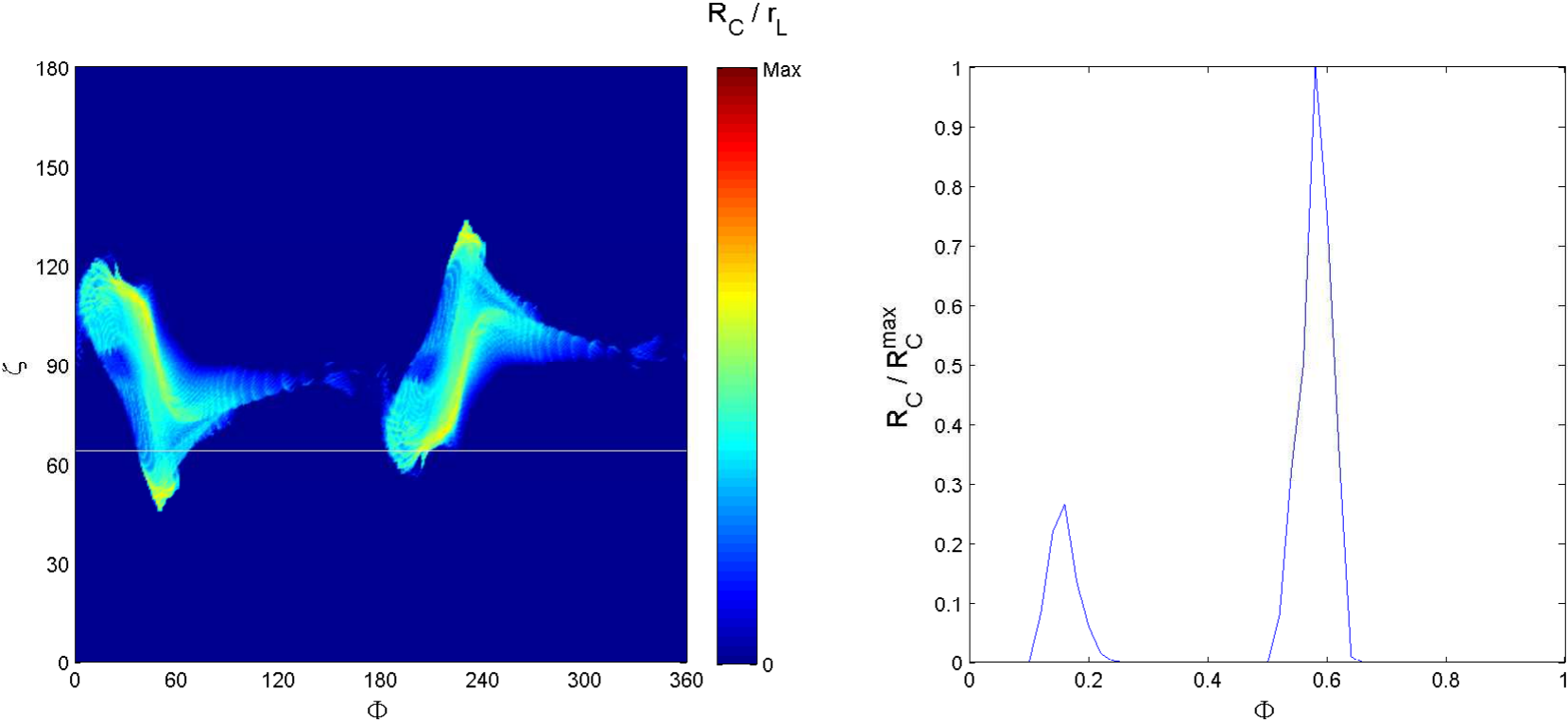}
\end{tabular}
\caption{The sky map of $R_{\rm C}$  and the variation of $R_{\rm C}$ with the pulsed phase for the same model parameters as in Figure \ref{Fig2}. \\}
\label{Fig5}
\end{figure*}

\section{curvature radiation}
The particle velocity in the dissipative magnetosphere is defined by \citep{gru12,pet20b,cai23}
\begin{eqnarray}
{\bf v_{\pm}}=  {{\bf E} \times {\bf B}\pm(B_0{\bf {B}}+E_0{\bf {E}}) \over B^2+E^2_{0}},
\label{Eq6}
\end{eqnarray}
where the two signs correspond to positrons and  electrons, they follow a differen trajectory that  only depends on the local electric and
magnetic field. The Lorentz factor along particle trajectory is integrated by
\begin{eqnarray}
\frac{d\gamma}{dt}=\frac{q_{\rm e}c E_{\rm acc}}{m_{\rm e}c^2}- \frac{2q^2_{\rm e} \gamma^4}{3R^2_{\rm CR}m_{\rm e}c} ,
\end{eqnarray}
which takes into account the influence of the local accelerating electric field and the curvature radiation losses.
The  photon spectrum of the particle curvature radiation  with Lorentz factor $\gamma$ is given by
\begin{eqnarray}
F(E_{\gamma},r)=\frac{\sqrt{3} e^2 \gamma}{2 \pi \hbar R_{\rm C} E_{\gamma}}F(x)\;,
\end{eqnarray}
where $R_{\rm C}$ is the curvature radius of particles, $x=E_{\gamma}/E_{\rm cur}$, $E_{\gamma}$ is the  radiation photon energy, $E_{\rm cur}=\frac{3}{2}c\hbar\frac{\gamma^3}{R_{\rm CR}}$ is the characteristic energy of the curvature radiation photon,  and the function $F(x)$ is given by
\begin{equation}
F(x)=x\int_{x}^{\infty}{K_{\rm 5/3}}(\xi)\;d\xi,
\end{equation}

The  particles are randomly  injected at the stellar surface with the small Lorentz factor. We integrate the particle trajectory from the the neutron star surface up to $r=2.5 \, r_{\rm L}$  by equation (8). The Lorentz factor along each trajectory is then computed by equation (9). We assume the direction of the photon emission along the direction of particle motion, the direction of the photon emission and the  curvature radiation photons are then computed along each trajectory.  The sky maps are produced by collecting  all curvature radiation photons  in the ($\zeta$,$\phi$)
plane.  The energy-dependent light curves are obtained by accumulating all emitting photons at a given energy range for a  constant viewing angle $\zeta$. The phase-averaged spectra are  obtained by
averaging  all  emitting photons in energy over the observed phase $\phi$ for a constant $\zeta$. The phase-resolved spectra are obtained by collecting all emitting photons in energy at a given $\phi$ range for a  constant $\zeta$.\\

\section{results}
The  light curves can provide an  diagnostic for pulsar magnetosphere geometry, while the  energy spectra can both  explore the  geometry  and location of the emission regions in the  magnetosphere. The combined light curves and spectra can put a stronger constraint on the  location of the particle acceleration and the geometry of the emission regions in the  magnetosphere. We simultaneously fit the  energy-dependent light curves, phase-averaged spectra and phase-resolved spectra of  the Vela pulsar to constrain the model parameters. We show the sky maps and the predicted $\gamma$-ray light curves at $>0.1$ GeV energies for the Vela pulsar in Figure \ref{Fig2}. We can see that the predicted sky maps show the two bright caustics that are characteristic of
radiation from the current sheet. Our sky maps show that the emission pattern from the current sheet is non-uniform, which depends on the field line geometry and the $E_{\rm acc}$ distribution in the current sheet. On the one hand, the emission from the current sheet is concentrated in the same region of the sky map due to the ``stagnation" caustic effects in the near force-free magnetosphere \citep{bai10}. On the other hand, the  $E_{\rm acc}$ distribution in the current sheet is non-uniform, the $E_{\rm acc}$ is more confined to  the current sheet in the range of  1$-$2\,$R_{\rm L}$, and the $E_{\rm acc}$  varys with radius and azimuth and become very weak beyond the $2\,R_{\rm L}$. Therefore, we can not see an extended geometric emission pattern with $\pm \, 65^{\circ}$ above/below the equatorial plane in sky map due to the non-uniform emission from the current sheet. In fact, our emission  pattern from the current sheet is also very similar to those of the PIC model \citep{phi18,kal18}.

In our model, the field line geometry and the $E_{\rm acc}$ distribution determines the light curve shape, while the $E_{\rm acc}$ value determines the peak energy of the $\gamma$-ray spectrum, which can be adjusted by the free parameter $\kappa$. The free parameter $\kappa$ is chosen to adjust the $E_{\rm acc}$ value to match the  Fermi $\gamma$-ray spectra with the real pulsar parameters. The model parameters are  $B_{*}=4\times10^{12} \, \rm G$, $P=0.089 \, \rm s$, $\kappa=17$, $\alpha=65^\circ$ and $\zeta=64^\circ$.  The parameter $\kappa=17$ is only chosen to obtain the suitable $E_{\rm acc}$ value to match the Fermi $\gamma$-ray spectra with the real Vela pulsar parameters of $B_{*}=4\times10^{12} \, \rm G$ and $P=0.089 \, \rm s$. The same results can also be obtained by using a simulated  pair multiplicity $\kappa=3$ and a slightly lower surface magnetic field $B_{*}=7\times10^{11} \, \rm G$. The pair multiplicity $\kappa=17$  can not be understood as the real Veal pulsar parameter in our model, which is only physically related to the pair cascade efficiency in ``killing" $E_{\rm acc}$.
We find that the predicted light curve  can well match the Fermi observed $\gamma$-ray data for the Vela pulsar. The peak phase and the  peak separation of the Vela pulsar can be well reproduced by our model. Our model can better match the $\gamma$-ray light curves of the Vela pulsar compared to those in \citet{har21} and \citet{bar22}. They can not  provide a good match to the peaks phases of the Vela pulsar and the predicted first peak phase is larger than the observed one. We show the predicted energy-dependent $\gamma$-ray light curves  and compare them with those of the Vela pulsar observed by Fermi  in Figure \ref{Fig3}.  We observe that  the flux of the first peak relative to the second one decreases towards  the higher photon energies  and the peak width also narrow with increasing photon energies.  Our model can well reproduce the decreasing ratio of the first peak to the second one ($\rm P1/P2$) and the narrow of the peak width. Our results indicate that the highest-energy photons originate from the second peak of the Fermi light curves. The predicted general trend of the light curves with the energy evolution is in good agreement with the Fermi observed one. We show the predicted phase-averaged spectra and the phase-resolved spectra of the first and second peaks for the Vela pulsar in Figure \ref{Fig4}. We see that the Fermi observed phase-averaged spectra can well be explained by the curvature radiation from the current sheet outside the LC. Our model also can well reproduce the Fermi observed phase-resolved spectra of the first and second peaks for the Vela pulsar. We find that the phase-resolved spectra of the second peak  have a  higher flux and a  larger spectral cutoff compared to the first peak one. This is because that the light curves of the second peak survive at the higher energy compared to the first peak one. We show the sky map of $R_{\rm C}$ and the variation of $R_{\rm C}$ with the pulsed phase in Figure \ref{Fig5}. We see that the sky pattern of $R_{\rm C}$ is similar to that of the caustic emission, because the variation of $R_{\rm C}$  closely reflects the variations of the caustic emission. We also find that the curvature radius in the second peak is the systematically larger than that in the first peak. The  cutoff energy $E_{\rm cut}$ of the curvature radiation in the radiation reaction limit  is related with the accelerating electric field  and the curvature radius  by
\begin{eqnarray}\label{eqn-Ec}
E_{\rm cut}\propto\,E^{3/4}_{\rm acc}R^{1/2}_{\rm C},
\end{eqnarray}
which scales with $R^{1/2}_{\rm C}$. Therefore, we can expect a large $E_{\rm cut}$ value in the second peak due to the systematically larger curvature radius than that in the first peak. Therefore,  the decreasing $\rm P1/P2$ ratio with increasing energy can be explained by  the systematically larger curvature radius of the second  peak than the first one. A similar result is also found in the force-free magnetosphere by  \citet{har21} and \citet{bar22}.

\section{Discussion and Conclusions}
We  explore  the properties of the pulsar energy-dependent $\gamma$-ray light curves and  spectra in the dissipative magnetospheres. The dissipative magnetospheres with the combined FFE and AE are
constructed by  applying the force-free description where $E \leq B$ and the AE description where $E > B$.
We compute the dissipative magnetospheres by a pseudo-spectral method with the high-resolution simulations in the rotating coordinate system. Our simulations show  that the dissipative magnetospheres have the near force-free field structure and the dissipative region is confined to only near the equatorial current sheet outside the LC. We compute the energy-dependent $\gamma$-ray  light curves, phase-average and phase-resolved spectra by the test particle trajectory method in the dissipative magnetosphere, taking into account both the local accelerating electric field and  radiation reaction. We can generally reproduce the Fermi observed energy-dependent $\gamma$-ray  light curves and spectra of the Vela pulsar. The decreasing ratio of the first peak to the second one  with increasing energy can well reproduced by our model, which is attributed to the systematically larger local curvature radius and the higher spectral cutoff in the second peak.

\cite{bra15} studied  the energy-dependent $\gamma$-ray  light curves of the Vela pulsar by the FIDO (Force-free Inside Dissipative Outside) model. They can reproduce the observed trends of the light curve with the energy evolution for the Vela pulsar. However, an approximate accelerating electric field is used based on the corresponding force-free solutions in their study, which is not self-consistently determined by the simulation themselves. Moreover, the predicted phase-averaged and phase-resolved spectra is  not directly compared with the observed ones. The energy-dependent light curves and spectra of the Vela pulsar is also explained by an extended slot gap and current-sheet model with the constant accelerating electric field in the force-free magnetospheres \citep{har21,bar22}. However, the predicted light curve with  the constant accelerating electric field  can not well match the  peaks phases of the Vela pulsar, because the accelerating electric field is the function of the altitude and azimuth. Our results  can  provide a better match to the peaks phases of the Vela pulsar by using the  accelerating electric field from the simulation themselves. Our MHD simulations show that the acceleration electric fields in the current sheet decrease with increasing pair multiplicity. However, the particle accelerations in the current sheet not only depend on the plasma supply from the pair multiplicity but also the reconnection rate in the current sheet. Therefore, it is necessary to further study magnetic reconnection acceleration in the current sheet by the PIC method.
The observed bridge emission of the Vela pulsar can not be explained by our model. It is suggested that the the bridge emission may originate in the inner regions of the magnetosphere \citep{bar22}. We will further explore the the bridge emission of the Vela pulsar by introducing the local dissipation not only in the current sheet region but also in the inner region. The study in the Fermi $\gamma$-ray band is not enough to distinguish between different radiation mechanisms and radiation locations in the pulsar magnetosphere. A future study on the multi-wavelength light curves and spectra can better constrain the radiation mechanisms and localize the  photon production sites in the magnetosphere. We will use the presented dissipative magnetospheres to present the study of the multi-wavelength light curves and spectra in a forthcoming work.

\acknowledgments
We thank the anonymous referee for valuable comments and suggestions.
We would like to thank  J\'{e}r$\hat{\rm o}$me P\'{e}tri for some useful discussions.  We acknowledge the financial support from the National Natural Science Foundation of China 12003026 and 12373045, and the Basic research Program of Yunnan Province 202001AU070070 and 202301AU070082.



\begin{thebibliography}{}
\bibitem[Abdo et al. (2010)]{abd10} Abdo, A. A., Ackermann, M., Ajello, M., et al. 2010, ApJS, 187, 460
\bibitem[Abdo et al. (2013)]{abd13} Abdo, A. A., Ajello, M., Allafort, A., et al. 2013, ApJS, 208, 17
\bibitem[Bai \& Spitkovsky (2010)]{bai10} Bai, X.N., \& Spitkovsky, A. 2010, ApJ, 715, 1282
\bibitem[Barnard et al. (2022)]{bar22} Barnard, M., Venter, C., Harding, A. K., Kalapotharakos, C., \& Johnson, T. 2022, ApJ, 925, 184
\bibitem[Belyaev (2015)]{bel15} Belyaev, M. A. 2015, MNRAS, 449, 2759
\bibitem[Brambilla et al. (2015)]{bra15} Brambilla, G., Harding, A. K., Kalapotharakos, K., \& Kazanas, D. 2015, ApJ, 804, 84
\bibitem[Brambilla et al. (2018)]{bra18} Brambilla, G., Kalapotharakos, K., Timokhin, A. N., Harding, A. K., \& Kazanas, D. 2018, ApJ, 858, 81
\bibitem[Bogovalov et al. (2018)]{bog18} Bogovalov, S. V., Contopoulos, I., Prosekin, A., Tronin, I., \& Aharonian, F. A. 2018, MNRAS, 476, 4213
\bibitem[Cai et al. (2023)]{cai23} Cai, Y., Gralla, S. E., \& Paschalidis, V. 2023, PhRvD, 108, 063019
\bibitem[Cao et al. (2016a)]{cao16a} Cao, G., Zhang, L., \& Sun, S. N. 2016a, MNRAS, 455, 4267.
\bibitem[Cao et al. (2016b)]{cao16b} Cao, G., Zhang, L., \& Sun, S. N. 2016b, MNRAS, 461, 1068.
\bibitem[Cao \& Yang (2019)]{cao19} Cao, G., \& Yang, X. B. 2019, ApJ, 874, 166
\bibitem[Cao \& Yang (2020)]{cao20} Cao, G., \& Yang, X. B. 2020, ApJ, 889, 29
\bibitem[Cao \& Yang (2022)]{cao22} Cao, G., \& Yang, X. B, 2022, ApJ, 925, 130
\bibitem[Chen \& Beloborodov (2014)]{che14} Chen, A. Y., \& Beloborodov A. M. 2014, ApJ, 795, L22

\bibitem[Contopoulos et al. (1999, hereafter CKF)]{con99} Contopoulos, I., Kazanas, D., \& Fendt, C. 1999, ApJ, 511, 351
\bibitem[Contopoulos \& Kalapotharakos (2010)]{con10} Contopoulos, I., \& Kalapotharakos, C. 2010, MNRAS, 404, 767
\bibitem[Contopoulos et al. (2016)]{con16} Contopoulos I. 2016, \mnras, 463, L94

\bibitem[Cerutti et al. (2015)]{cer15} Cerutti, B., Philippov, A., Parfrey, K., \& Spitkovsky, A. 2015, MNRAS, 448, 606
\bibitem[Cerutti et al. (2016)]{cer16} Cerutti B., Philippov A. A., \& Spitkovsky, A. 2016, MNRAS, 457, 2401
\bibitem[Goldreich \& Julian (1969)]{gol69} Goldreich, P., \& Julian, W. H. 1969, ApJ, 157, 869


\bibitem[Gruzinov (2012)]{gru12} Gruzinov, A. 2012, arXiv: 1205.3367
\bibitem[Gruzinov (2013)]{gru13} Gruzinov, A. 2013, arXiv:1303.4094
\bibitem[Harding \& Kalapotharakos (2015)]{har15} Harding, A. K., \& Kalapotharakos, C. 2015, ApJ, 811, 63
\bibitem[Harding et al. (2018)]{har18} Harding, A. K., Kalapotharakos, C., Barnard, M., \& Venter, C. 2018, ApJ, 869, L18
\bibitem[Harding et al. (2021)]{har21} Harding, A. K., Venter, C., \& Kalapotharakos, C. 2021, ApJ, 923, 194
\bibitem[Kalapotharakos \& Contopoulos (2009)]{kal09} Kalapotharakos, C., \& Contopoulos, I. 2009, A\&A, 496, 495
\bibitem[Kalapotharakos et al. (2012a)]{kal12a} Kalapotharakos, C., Kazanas D., Harding, A., \& Contopoulos, I. 2012a, ApJ, 749, 2
\bibitem[Kalapotharakos et al. (2012b)]{kal12b} Kalapotharakos, C., Harding, A. K., Kazanas, D., \& Contopoulos, I. 2012b, ApJL, 754, L1
\bibitem[Kalapotharakos et al. (2014)]{kal14} Kalapotharakos, C., Harding, A. K., \& Kazanas, D. 2014, ApJ, 793, 97
\bibitem[Kalapotharakos et al. (2017)]{kal17} Kalapotharakos, C., Harding, A. K., Kazanas, D., \& Brambilla, G. 2017, ApJ, 842, 80
\bibitem[Kalapotharakos et al. (2018)]{kal18} Kalapotharakos, C., Brambilla, G., Timokhin, A., Harding, A. K., \& Kazanas, D. 2018, ApJ, 857, 44

\bibitem[Li et al. (2012)]{li12} Li, J., Spitkovsky, A., \& Tchekhovskoy, A. 2012, ApJ, 746, 60
\bibitem[Muslimov \& Harding (2005)]{mus05} Muslimov, A. G., \& Harding, A. K. 2005, ApJ, 630, 454
\bibitem[Philippov \& Spitkovsky (2014)]{phi14} Philippov, A. A., \& Spitkovsky, A. 2014, ApJ, 785, L33
\bibitem[Philippov et al. (2015)]{phi15} Philippov, A. A., Spitkovsky, A., \& Cerutti, B. 2015, ApJ, 801, L19
\bibitem[Philippov \& Spitkovsky (2018)]{phi18} Philippov, A. A., \& Spitkovsky, A. 2018, ApJ, 855, 94

\bibitem[Parfrey et al. (2012)]{par12} Parfrey K., Beloborodov A. M., \& Hui L. 2012, MNRAS, 423, 1416
\bibitem[P\'{e}tri (2012)]{pet12} P\'{e}tri, J. 2012, MNRAS, 424, 605
\bibitem[P\'{e}tri (2016)]{pet16} P\'{e}tri, J. 2016, MNRAS, 455, 3779
\bibitem[P\'{e}tri (2020a)]{pet20a} P\'{e}tri, J. 2020a, MNRAS, 491, 46
\bibitem[P\'{e}tri (2020b)]{pet20b} P\'{e}tri, J. 2020b, Universe, 6, 15
\bibitem[P\'{e}tri (2022)]{pet22} P\'{e}tri, J. 2022, MNRAS, 512, 2854
\bibitem[Spitkovsky (2006)]{spi06} Spitkovsky, A. 2006, ApJ, 648, L51
\bibitem[Yang \& Cao (2021)]{yang21} Yang, X. B., \& Cao, G. 2021, ApJ, 909, 88

\end{thebibliography}
\end{document}